\documentstyle[eqsecnum,aps,graphicx]{revtex}

\def\nim{Nucl.\ Inst.\ Meth.\ } 
\begin{document}
\pretolerance=10000

\title{\bf Observation of the Shadowing of Cosmic
Rays by the Moon using a Deep Underground Detector}

\author{
\pretolerance=10000
 \begin{center}
{\bf The MACRO Collaboration}\\
\nobreak\bigskip\nobreak\footnotesize
M.~Ambrosio$^{12}$, 
R.~Antolini$^{7}$, 
C.~Aramo$^{7,n}$,
G.~Auriemma$^{14,a}$, 
A.~Baldini$^{13}$, 
G.~C.~Barbarino$^{12}$, 
B.~C.~Barish$^{4}$, 
G.~Battistoni$^{6,b}$, 
R.~Bellotti$^{1}$, 
C.~Bemporad $^{13}$
% ...
\\
\vspace{1.5 cm}
\footnotesize
1. Dipartimento di Fisica dell'Universit\`a di Bari and INFN, 70126 
Bari,  Italy \\
2. Dipartimento di Fisica dell'Universit\`a di Bologna and INFN, 
 40126 Bologna, Italy \\
3. Physics Department, Boston University, Boston, MA 02215, 
USA \\
4. California Institute of Technology, Pasadena, CA 91125, 
USA \\
5. Department of Physics, Drexel University, Philadelphia, 
PA 19104, USA \\
6. Laboratori Nazionali di Frascati dell'INFN, 00044 Frascati (Roma), 
Italy \\
7. Laboratori Nazionali del Gran Sasso dell'INFN, 67010 Assergi 
(L'Aquila),  Italy \\
8. Depts. of Physics and of Astronomy, Indiana University, 
Bloomington, IN 47405, USA \\
9. Dipartimento di Fisica dell'Universit\`a dell'Aquila  and INFN, 
 67100 L'Aquila,  Italy \\
10. Dipartimento di Fisica dell'Universit\`a di Lecce and INFN, 
 73100 Lecce,  Italy \\
11. Department of Physics, University of Michigan, Ann Arbor, 
MI 48109, USA \\	
12. Dipartimento di Fisica dell'Universit\`a di Napoli and INFN, 
 80125 Napoli,  Italy \\	
13. Dipartimento di Fisica dell'Universit\`a di Pisa and INFN, 
56010 Pisa,  Italy \\	
14. Dipartimento di Fisica dell'Universit\`a di Roma ``La Sapienza" and INFN, 
 00185 Roma,   Italy \\ 	
15. Physics Department, Texas A\&M University, College Station, 
TX 77843, USA \\	
16. Dipartimento di Fisica Sperimentale dell'Universit\`a di Torino and INFN,
 10125 Torino,  Italy \\	
17. Also  Faculty of Sciences, University Mohamed I, B.P. 424 Oujda, Morocco \\
$a$ Also Universit\`a della Basilicata, 85100 Potenza,  Italy \\
$b$ Also INFN Milano, 20133 Milano, Italy\\
$c$ Also Istituto TESRE/CNR, 40129 Bologna, Italy \\
$d$ Also Scuola Normale Superiore di Pisa, 56010 Pisa, Italy\\
$e$ Also Dipartimento di Fisica, Universit\`a di Roma Tre, Roma, Italy \\
$f$ Also Universit\`a di Trieste and INFN, 34100 Trieste, 
Italy \\
$g$ Also Dipartimento di Energetica, Universit\`a di Roma, 
 00185 Roma,  Italy \\
$h$ Also Institute for Nuclear Research, Russian Academy
of Science, 117312 Moscow, Russia \\
$i$ Also Institute for Space Sciences, 76900 Bucharest, Romania \\
$l$ Swarthmore College, Swarthmore, PA 19081, USA\\
$m$ RPD, PINSTECH, P.O. Nilore, Islamabad, Pakistan \\
$n$ Also INFN Catania, 95129 Catania, Italy\\
$o$ Also Department of Physics, Pennsylvania State University, 
University Park, PA 16801, USA\\
\end{center}
}

\maketitle
\begin{abstract} 

Using data collected by the MACRO experiment during the years 1989-1996, we
show evidence for the shadow of the moon in the underground cosmic ray flux
with a significance of 3.6$\sigma$.  This detection of the shadowing effect
is the first by an underground detector.  A maximum-likelihood analysis is
used to determine that the angular resolution of the apparatus is $0.9^\circ
\pm 0.3^\circ $.  These results demonstrate MACRO's capabilities as a muon
telescope by confirming its absolute pointing ability and quantifying its
angular resolution.

\end{abstract}

\vspace{0.75in}

\section{Introduction} 

MACRO is a large area underground detector located in Hall~B of the Gran
Sasso National Laboratory (LNGS) in Italy at an average depth of
3700~m.w.e.  
The full apparatus has dimensions 76.5m$\times$12m$\times$9.6m.
The detector's active
technologies include liquid scintillation counters and streamer tubes.  
Low-radiation crushed rock separates 
the active detector planes.  Normally incident muons from above lose $\approx
 1.5$ GeV in  traversing the detector; at the surface, throughgoing muons have
 energies in excess of $\sim 3$~TeV, corresponding to 
primary cosmic rays of  energy $>~7$~TeV.  The 
average energy of the muons detected by
MACRO is $\sim 300$ GeV.  
A detailed description of the 
detector can be found in \cite{MACRO}.  

MACRO was primarily designed to
search for monopoles and rare particles in the cosmic rays, including
high energy neutrinos and muons from cosmic point sources \cite{MACRO}. 
 Since high energy neutrinos are identified as upward-going muon
 secondaries from neutrino interactions in the rock below the detector, 
MACRO functions as a muon telescope in  its search for muons and neutrinos 
from cosmic sources.  The purpose of this investigation is to  
verify its absolute pointing 
and quantify its angular resolution to point sources.  
Confidence in pointing is clearly essential for a telescope, and 
a preliminary determination of MACRO's pointing has been made by
bootstrapping highway surveys from the surface into the tunnel.  
However, a check of the accuracy of this determination is important. 
Further, reconstructed muon tracks
point back to an area on the sky whose width
depends on the intrinsic angular resolution of the detector as well as 
multiple Coulomb scattering in the rock overburden.  Consequently, the 
quality of MACRO as a muon telescope depends on its angular resolution.
As the required search region around a source shrinks with improving 
resolution, the contribution of events from the flat or slowly 
varying background also decreases relative to the signal.  

Traditional astronomical telescopes use observable point sources to
determine pointing and angular resolution.  Despite early reports of 
their existence \cite{mfm}, however, there are as yet no
established cosmic muon sources.   
In the absence of cosmic sources, we adopt the approach of the
CYGNUS \cite{cygnus}, CASA \cite{casa}, and Tibet \cite{tibet} air
shower arrays who used the moon as a fiducial object.  By blocking 
cosmic ray primaries, the moon appears as
a cosmic ray antisource or ``shadow'' \cite{clark} which can be 
used to verify MACRO's absolute pointing and to determine its angular 
resolution.

\section{The muon data sample and the expected background}

\subsection{Muon data sample}

MACRO consists of six nearly identical units called
supermodules, each 
of dimension 12.6m$\times$12m$\times$9.6m. 
The lower 4.8m of each supermodule consists of 10
horizontal planes of streamer tubes of dimension 12m$\times$6m. 
The 8 innermost 
planes are separated by seven layers of 60 gm/cm$^2$ absorber. 
The two outermost planes are each 
separated from the next nearest streamer tube plane by a 25 cm layer of
liquid scintillator.  The lateral walls consist
of stacked tanks of liquid scintillator, 25 cm thick,
sandwiched between six vertical streamer tubes planes. 
The upper 4.8~m of each supermodule (the attico) is a hollowed out 
version of the lower supermodule.  It 
consists of a top plane of liquid scintillator 
with two streamer tube planes on both above and below, and 
lateral walls 
of liquid scintillator 
with three vertical streamer tubes planes outside and inside.
All streamer tube wires are read out, providing the X coordinate on the
horizontal planes and the Z coordinate on the vertical planes. On the
horizontal planes the second coordinate, D, is obtained by 
horizontal aluminum strips oriented $26.5^0$
with respect to the streamer tube axis. These strips allow stereoscopic
reconstruction.  
For a complete description of the detector, see \cite{MACRO}.  

The muon sample used for the present analysis includes all events collected
from the start of MACRO data taking in February, 1989 through the end of
1996. The sample totals $39.3 \times 10^6$ events collected over $2.7
\times 10^3$ live~days.  
During the first part of this period the apparatus was under
construction.  Long running periods included one supermodule without
the attico ($A_{eff}\Omega \approx 1,010$~m$^2$~sr), six supermodules
without attico ($A_{eff}\Omega \approx 5,600$~m$^2$~sr) and finally the full
six supermodules with attico ($A_{eff}\Omega \approx 6,600$~m$^2$~sr).
Approximately
60\% of the data sample was obtained during periods when MACRO had full
acceptance.

The criteria used to select events for this analysis were designed to
optimize the quality of reconstructed tracks.  The selected events are
consequently those which most accurately point back to their origin on
the celestial sphere.  The specific data cuts used in this analysis are
listed below.

\begin{itemize}

   \item \underline{Run cuts}

       \begin{enumerate}
         \item Runs less than one hour in length were cut since short 
           runs were usually abnormally terminated and often 
           contained malfunctioning hardware or software.
         \item Runs with large numbers of UT clock errors were cut.
       \end{enumerate}

   \item \underline{Event cuts}

      \begin{enumerate}
         \item Only single and double muon events were retained; events with
                    multiplicity $> 2$ were cut.  The more complicated task 
                 of reconstructing multimuon tracks is more likely to 
                 introduce tracking errors.
         \item Successful track reconstruction requires a minimum of
           four crossed horizontal streamer tube planes; events with 
           fewer than four
           crossed horizontal planes were cut.
         \item Events crossing fewer than 3 streamer tube planes in
           the lower supermodules were cut.  This cut removes low energy
           events that pass through the attico without crossing any rock
           absorber layers.  Such events have large
           multiple Coulomb scattering angles and so are spread widely
           with respect to their point of origin on the celestial
           sphere.
         \item Reconstructed tracks with 
           $\chi^{2}/d.o.f. > 1.5$ were cut.  This cut removes events
           with poorly reconstruced tracks as well as events with large
           numbers of hits outside the track.
         \item Events with different reconstructed
           multiplicity in the two streamer tube views were cut.
        \item Events with UTC clock errors were cut since the correct 
      time is necessary to project the track back onto the celestial sphere.
       \end{enumerate}
 
\end{itemize}

\noindent These selection criteria reduce the sample size to 
$30.51 \times 10^6$ muons.

The topocentric position of the moon was computed at the arrival time
of each event in the sample using the database of ephemerides available
from the Jet Propulsion Laboratory, JPLEPH \cite{jpl}.  A correction for
the parallax due to MACRO's instantaneous position on the earth was
applied to each ephemeris position \cite{duffet}.  Since the parallax
correction requires an accurate computation of the earth-moon distance,
this calculation also results in an accurate determination of angular
size subtended by the moon at MACRO.

The muon events in a window $10^\circ$ on a side and centered on the moon 
were selected for further analysis.  There are $2.3 \times 10^5$ events that
pass all cuts in this window.


\subsection{Expected background}

Twenty five background samples were generated for each run used in the
analysis.  These backgrounds were constructed by coupling the direction
of each muon in the run with the times of 25 randomly selected muons
from the same run.  The 25 background samples were then processed using 
the same procedure as the muon data sample.

\section{Shadow of the Moon}

\subsection{Event deficit around the moon}

\begin{figure}[t]
\begin{center}
\includegraphics[height=8.6cm,width=8.6cm]{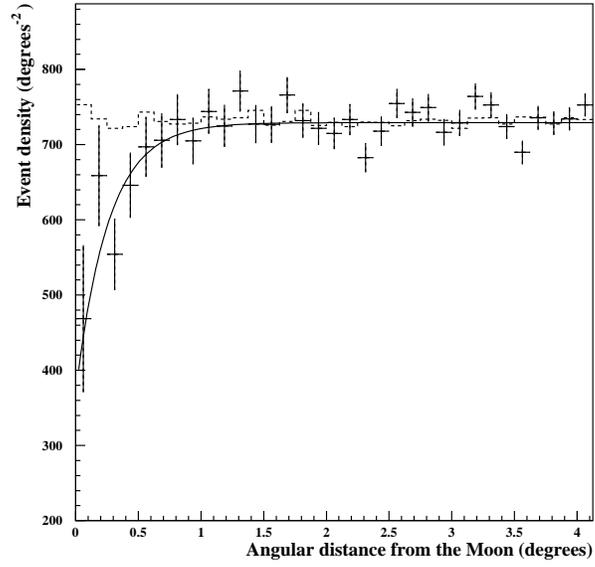}
\caption{The event density vs angular distance from the moon 
  center in bins of equal angular width.  The width of each bin 
is   $0.125^\circ$.  The dashed curve is the average
  expected background computed from 25 background samples.  The solid
  curve shows the expected event density as computed for an angular
  resolution of the MACRO apparatus of $0.90^\circ$.}
\label{moondens}
\end{center}
\end{figure}

Each muon event in the window was sorted into bins of equal angular width as
a function of angular distance from the moon center.  The angular width of
each bin was $0.125^\circ$, which gives a solid angle for the $i_{th}$ bin of
$\Delta \Omega_i = (2i - 1) \cdot 0.05$ deg$^2$.  Once filled, the contents
of each bin, $N_i$, were divided by the solid area of the bin, resulting in
the distribution, $\Delta N / \Delta \Omega$ which approximates the
differential event density as a function of angular distance from the moon
center.  Figure~\ref{moondens} shows as data points the computed differential
event density.  The errors on the data points are statistical.  Superposed on
this distribution is the expected event density in the
absence of the shadowing effect, $\Delta N^{bkd} / \Delta \Omega$.  This
background distribution was determined by averaging over the 25 background
samples the number of events in each solid angle bin, and then dividing the
result by the solid angle of the bin.  This nearly flat distribution is
described reasonably well by a constant event density $\approx 735$
deg$^{-2}$.  The shadowing effect, or the deviation of the angular event
density from the background, is clearly evident.  This figure confirms that
reconstructed muon tracks can be accurately pointed back to celestial
coordinates, thus confirming MACRO's absolute pointing ability.

We have made a simple estimate of the significance of this detection of 
the moon shadow using the information given in Figure~\ref{moondens}.  
First, we computed the integral event deficit as a function of the angular
 distance from the moon center.  Out to bin $n$, the integral event deficit, 
$\Delta_n  N^{def} $, is given by 

\begin{equation}
   \Delta_n N^{def} =  \sum_{i = 1}^n [ N^{bkd}_i - N_i ] .
\end{equation}

\begin{figure}[t]
\begin{center}
\includegraphics[height=8.6cm,width=8.6cm]{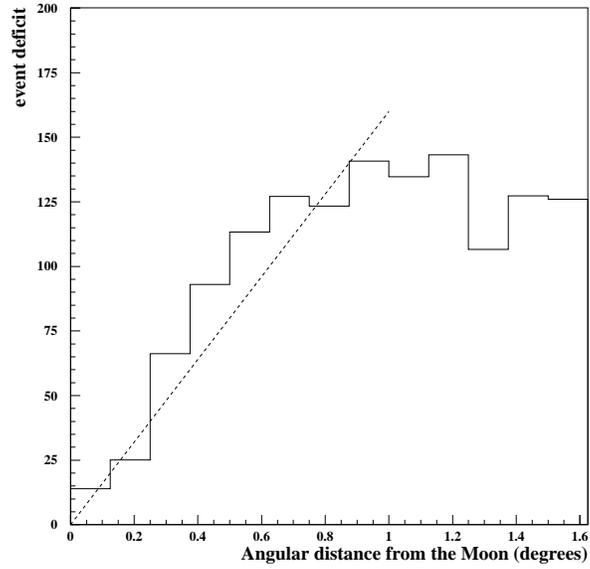}
\caption{Integral event deficit vs angular distance from
 the moon center.  Superposed on the integral event deficit 
is an approximate fit to the distribution out to $0.9^\circ$:
 $\Delta N^{def} = 165 \alpha$, where $\alpha$ is the angular distance from 
the moon center.}
\label{int_def}
\end{center}
\end{figure}

\noindent This distribution represents the cumulative number of events that 
the data distribution differs from the flat background and is shown in 
Figure~\ref{int_def}.  This figure shows that the deficit increases until 
the shadow can no longer be distinguished from the background at 
$\sim 0.9^\circ$.  Out to $0.9^\circ$ the integral deficit can be 
approximated by $\Delta N^{def} \approx 165 \alpha$, where $\alpha$ is 
the angular distance from the moon center; this simple approximation is 
shown as a dashed line in Figure~\ref{int_def}.  An estimate of the 
significance is then given by

\begin{equation}
  \frac{\Delta N^{def}}{\sqrt{N^{bkd}}}  = 
     \frac{165 \alpha}{\sqrt{735 \pi \alpha^2}} 
               \approx  3.5 .
\end{equation}

\noindent This computation shows that the moon shadow detection 
has a significance of approximately $3.5 \sigma$.  

\subsection{Maximum likelihood analysis}

In the simple deficit analysis above, we have implicitly assumed that the 
position of the moon's shadow is known and we have binned the events using
this information.  We now relax that 
assumption and search for the moon shadow in a
direction-independent way with the maximum likelihood method of
COS-B~\cite{COS-B}, a technique first described in detail by
Cash~\cite{cash}.  This method is based on {\it a priori} knowledge
of the point spread function of the MACRO detector (MPSF).

\begin{figure}[t]
\begin{center}

\includegraphics[height=8.6cm,width=8.6cm]{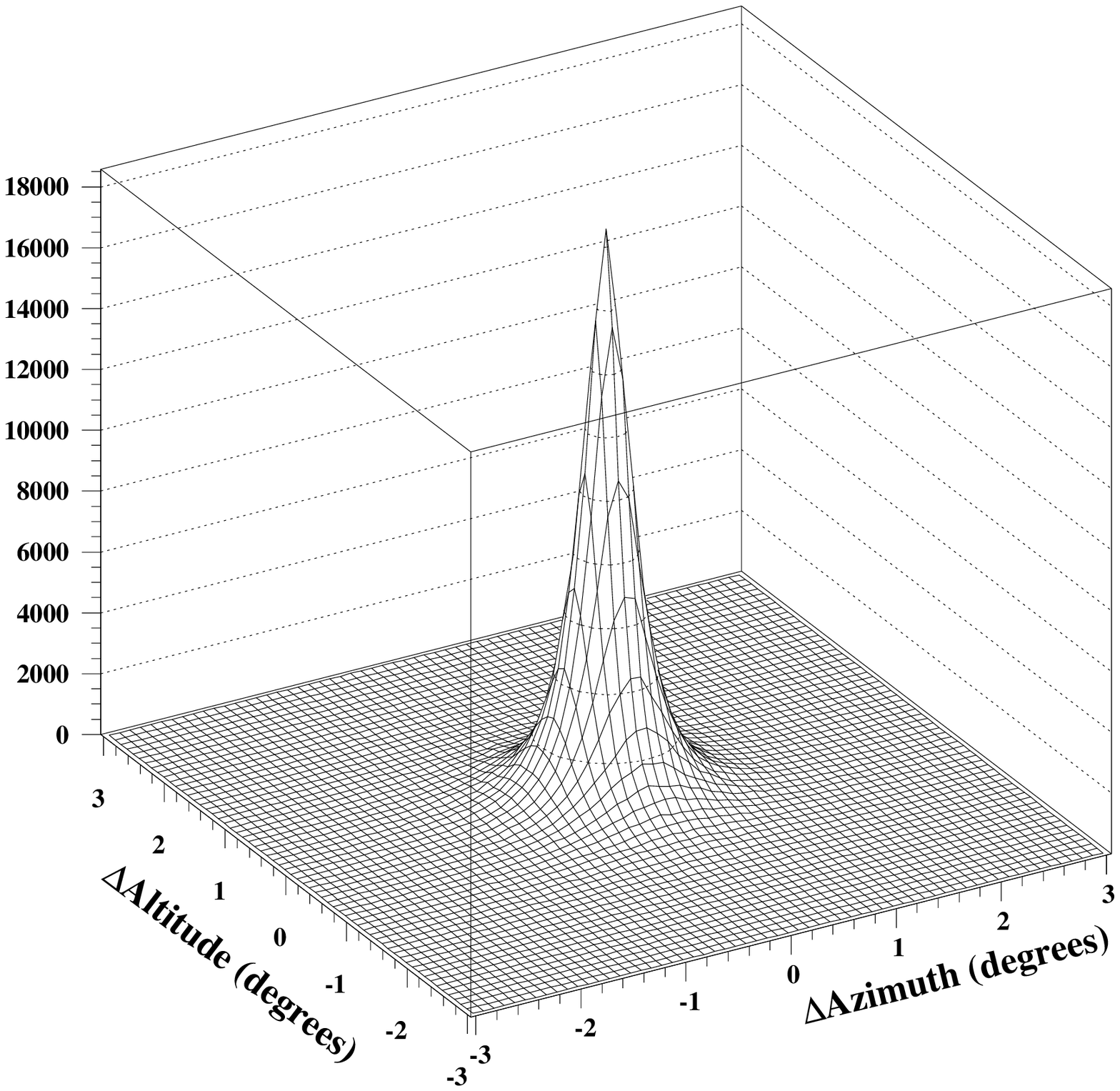}
\vspace{0.2cm}
\includegraphics[height=8.6cm,width=8.6cm]{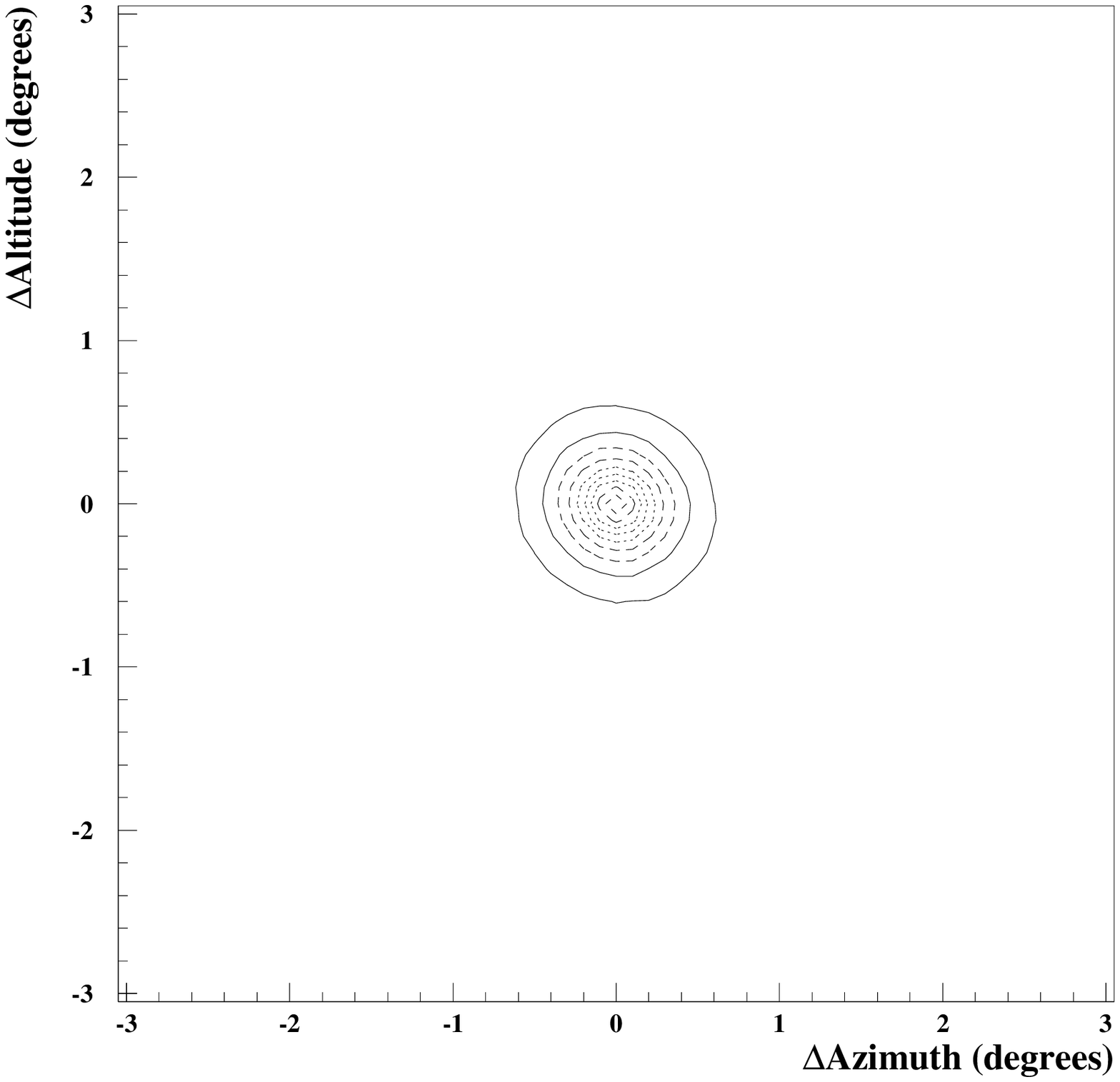}
\caption{(a) The point spread function of the MACRO apparatus (MPSF) derived
from the space angle distribution of 1,044,877 double muons.    
(b) Contour diagram of
  the MPSF.  The contours levels are equally
  spaced between 2,000 and 18,000 muons.  The MPSF is circularly
  symmetric in this coordinate system.}
\label{psf_altaz}
\end{center}
\end{figure}

We have determined the MPSF using the observed space angle
distribution of double muons.  As shown in \cite{multiple},
the number of double muons as a function of lateral separation
is a power law distribution that falls 
with a scale length of $\sim 15$~m. 
Muon pairs produced in a primary interaction 
at 20~km therefore have typical  
initial separation angles $\leq .05^\circ$. 
Since ``double'' muons
initially move along virtually parallel paths, the distribution of
their separation angles is a good measure of the deviations introduced
into their tracking parameters by both scattering in the mountain 
overburden and the detector's intrinsic angular resolution.  
The space angle distribution of ``double'' muons 
must be divided by $\sqrt{2}$ to obtain the 
MPSF since both muons deviate from their initial 
trajectories.  In
Figure~\ref{psf_altaz} the MPSF, as determined from 1,044,877
muon pairs, is shown in altitude and azimuthal coordinates.  
Figure~\ref{psf_altaz} shows the strongly peaked, non-Gaussian behavior
of the MPSF.  

\begin{figure}[t]
\begin{center}
\includegraphics[height=8.6cm,width=8.6cm]{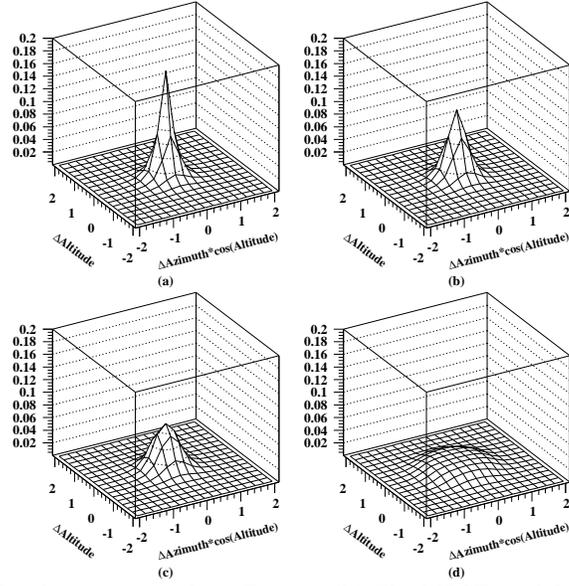}
\caption{
(a) The MPSF, normalized to unit area from Fig.~\ref{psf_altaz}a.  
(b) The MPSF modified by 
a disk of radius $0.26^\circ$, the average lunar radius, 
and then normalized to unit area; 
this is the modified MPSF used in the likelihood analysis.
(c) The MPSF modified by 
a disk of radius $0.4^\circ$ 
and then normalized. 
(d) The MPSF modified by 
a disk of radius $1.0^\circ$ 
and then normalized. 
}
\label{psf_moon} 
\end{center}
\end{figure}

To find the most likely position of the moon, we 
compare the two dimensional distribution of muons in the 
window centered on the moon with the expected background events in 
the same window.  In this analysis, 
each muon event is first sorted into a grid of equal solid angle
bins ($\Delta \Omega = 0.125^\circ \times 0.125^\circ = 1.6 \times
10^{-2}$ deg$^{2}$).  
The shadowing source of strength $S_M$ 
at fixed position $(x_s, y_s)$ 
that best
fits the data is then found by 
minimizing 

\begin{equation}
\chi^2 (x_s,y_s,S_M)=2 \sum_{i=1}^{n_{bin}}[ N_i^{ex}- N_i
  +N_i\ln{\frac{N_i}{N_i^{ex}}}],
\end{equation}

\noindent where the sum is over all bins in the window \cite{PDG}.  
Here $N_i$ is the number of events observed in each 
bin $i$, $N_i^{ex}$ is the expected number of events in bin $i$, and
$n_{bin}$ is the number of bins in the grid.  
This expression assumes that a Poissonian process
is responsible for the events seen in each bin.
The expected number of
events in bin $i$ is given by 

\begin{equation}
N_i^{ex}= N_i^{bkd} - S_M\cdot {\mathcal{P}} (x_s-x_i,y_s-y_i),
\end{equation}

\noindent where $N_i^{bkd}$ is the average number of background 
events at position 
$(x_i, y_i)$, and $S_M \cdot
{\mathcal{P}}(x_s-x_i,y_s-y_i)$ is the number of events removed from bin
$i$ by the shadow of the moon.  Here ${\mathcal{P}}(x_s-x_i,y_s-y_i)$ is the
MPSF, modified for the finite size of the moon's disk, 
computed at the point $(x_i,y_i)$ when the shadowing source is at
$(x_s,y_s)$.  The MPSF was modified by first selecting random 
positions on a disk with the average lunar radius, 0.26$^\circ$, and 
then drawing   
offsets from these positions 
from the MPSF distribution 
These new positions were rebinned into a new histogram and the
resulting distribution normalized to unit area.  
The modified point spread function is shown in Fig.~\ref{psf_moon}.
In this figure, the unmodified, normalized  
MPSF is shown in Fig.~\ref{psf_moon}a; 
the MPSF modified for the moon's finite 
disk is shown in Fig.~\ref{psf_moon}b.
In Figs.~\ref{psf_moon}c and d the MPSF has been modified by 
still larger shadowing disks.  The effect of the finite size
of the moon's disk does not have a large impact on 
the analysis.  

Finally, the shadow strength
$S_M$ that minimizes $\chi^2$ was computed  
for every grid point in the window.  
This 
minimum $\chi^2 (x_s,y_s,S_M)$ was then compared with 
$\chi^2(0)$ for the 
{\it null hypothesis} that no shadowing
source is present in the window ($S_M = 0$), 

\begin{equation}
  \lambda = \chi^2 (0) - \chi^2 (x_s,y_s,S_M).
\end{equation}

\noindent The most likely
position of the moon is the bin in which the maximum $\lambda$,
$\lambda^{max} \equiv \Lambda$, is found.  Since there is only one free
parameter, 
$S_M$, $\lambda$ behaves like $\chi^2_1$, a $\chi^2$ distribution with
one degree of freedom \cite{COS-B}.  The significance of the moon
detection is given by $\chi^2_1(\Lambda)$.

\begin{figure}[t]
\begin{center}
\includegraphics[height=8.6cm,width=8.6cm]{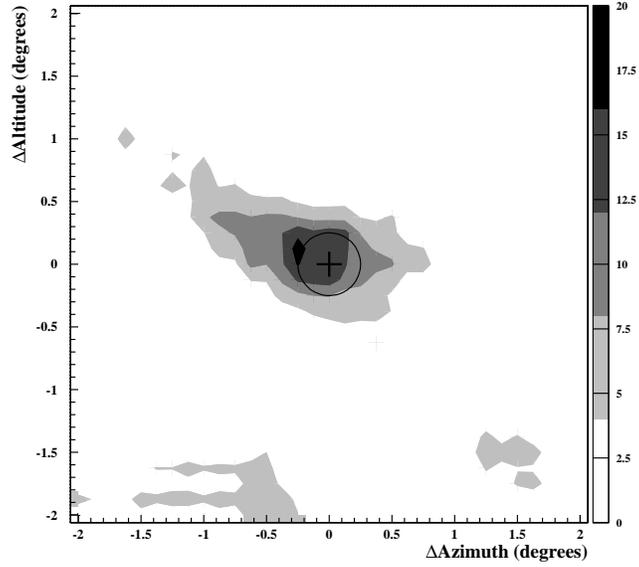}
\caption{The two dimensional distribution of $\lambda$ in bins of equal solid
angle in the moon window 
The axes are offsets from the moon center.
  The fiducial position of the moon, 
  at position (0,0), is marked by a $+$; a circle corresponding to the average
 lunar radius,
  $0.26^\circ$, is centered at this position.  The $\lambda$ grey scale
  is given at the right margin of the figure.  The maximum of this
  distribution, $\Lambda = 18.3$, is offset from the fiducial moon
  position at $\Delta \, {\rm Azimuth} = -0.25^\circ$ and $\Delta \, {\rm 
Altitude} = +0.125^\circ$.}
\label{lambda} 
\end{center}
\end{figure}

In Figure~\ref{lambda} we show the results of this analysis 
in a window $4.375^\circ \times
4.375^\circ$ centered on the moon.  This window has been 
divided into $35 \times 35$ cells, each having dimensions 
$0.125^\circ \times 0.125^\circ$.
In this figure, $\lambda$ is displayed in grey scale format
for every bin in the moon window. 
Also shown is the fiducial position of the moon  
and a circle centered at this position corresponding to the average
lunar radius, $0.26^\circ$.
The maximum  $\Lambda = 18.3$  
is found somewhat offset from the fiducial moon position at 
$\Delta \, {\rm Azimuth} = -0.25^\circ$, $\Delta \,
{\rm Altitude} = +0.125^\circ$ and   
provides further confirmation of
MACRO's absolute pointing.  
The value of  
the shadow strength 
at this position, $S_M = 153.9 \pm 37$~events, agrees well
with the observed value 155~events.  
Although the displacement from the fiducial position of 
the moon 
is not of high significance, we note 
that it is consistent with the deflection of 
cosmic ray primaries by the  geomagnetic field \cite{tibet}.
 
\begin{figure}[t]
\begin{center}
\includegraphics[height=8.6cm,width=8.6cm]{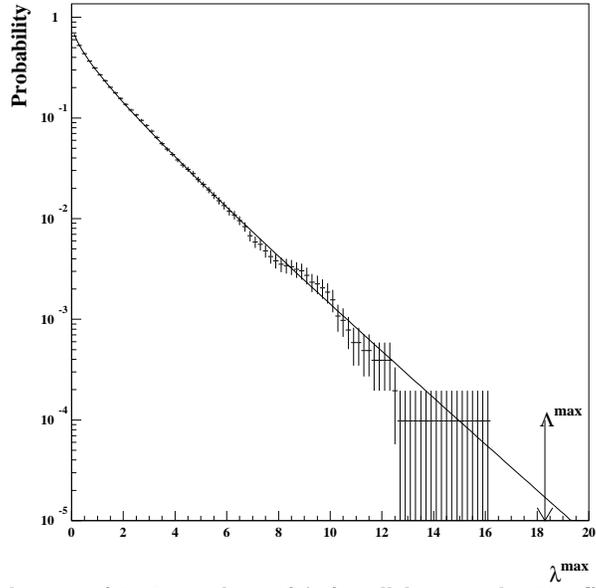}
\caption{The cumulative distribution of 10,224 values of 
  $\lambda$ for all bins in the 71 off-source windows.  
   Superposed on
  this distribution, as a solid curve, is the cumulative $\chi^2_1$
  distribution.  The arrow marks $\Lambda = 18.3$ from the likelihood analysis.}

\label{cum}
\end{center}
\end{figure}

We have verified the properties of the $\lambda$ distribution by
constructing 71 other windows similar to the moon window, each 
displaced from the next by $5^\circ$ in right ascension.  
For each off-source window, we followed the 
procedure used for the moon window in computing the expected
background.   To avoid edge 
effects associated with a source near the
edge of a window, we only evaluated $\lambda$ for the 
central 12$\times$12 bins.  In
Figure~\ref{cum} we have plotted the cumulative distribution of
the $12 \times 12 \times 71 = 10,224$ 
values of $\lambda$ for the 71 off-source
windows.  Superposed on this distribution as a solid curve is the 
cumulative $\chi^2_1$ distribution.  The distributions
agree well (the structure seen at $7 \leq \lambda \leq 10$ is due to 
roundoff error).  Thus, the probability of the detection of the
shadow of the moon at this bin location is $p(\chi^2_1 = 18.3) \leq 1.9
\times 10^{-5}$.  However, we would have considered the detection of the moon
shadow equally secure had the maximum been found at any of the 9 bins within
the geometric shadow, which reduces the probability of detection to $p \leq
1.7 \times 10^{-4}$.  The significance of the detection is therefore $3.6
\sigma$ and the results of the likelihood analysis and the deficit analysis
are in excellent agreement.   

As a crosscheck, a similar analyis
in right ascension and declination was performed.  The results were
equivalent, as required.   

\section{Angular resolution of the MACRO apparatus}

As shown in Figure~\ref{psf_altaz}, the MPSF cannot fit by a simple
Gaussian function.  Thus, the technique used 
by the CYGNUS \cite{cygnus},
CASA \cite{casa}, and Tibet \cite{tibet} air shower arrays, where such a
simplification was possible, must be modified.  
The air shower experiments  
find their PSF by rigidly scaling the dispersion of 
their Gaussian resolution functions and then computing the 
likelihood function for the detection of the moon shadow for each scaled value 
of $\sigma$.  The maximum likelihood  
defines the $\sigma$ to be 
used in the computation of the angular resolution.
In our approach, 
we first defined a scale parameter $\mathcal{F}$ that rigidly scales 
the modified MPSF by the factor $\mathcal{F}$,

\begin{displaymath}
  \tilde{\mathcal{P}}(x_s-x_i,y_s-y_i; \mathcal{F}).
\end{displaymath}

\noindent We then repeated the likelihood analysis in the moon 
window for different values of $\mathcal{F}$.  We assume that the value 
of $\mathcal{F}$ that maximizes $\Lambda$ gives the best $\tilde{\mathcal{P}}$ 
for computing the angular resolution.  

\begin{figure}[t]
\begin{center}
\includegraphics[height=8.6cm,width=8.6cm]{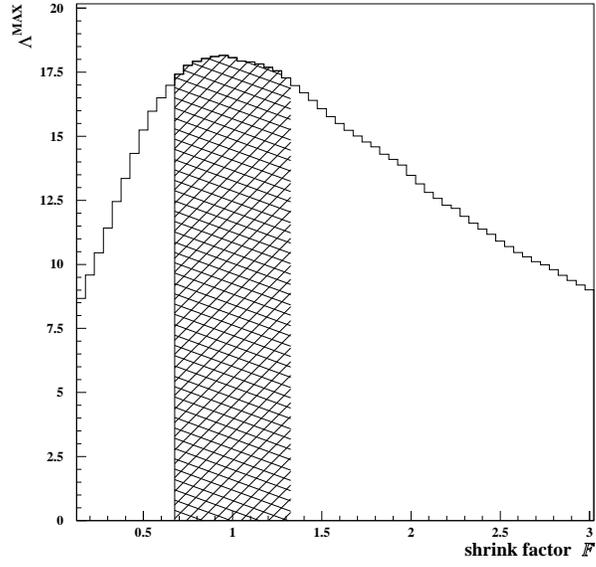}
\caption{The distribution of $\Lambda$ as a function of 
  $\mathcal{F}$.  The maximum $\Lambda$, $\Lambda^\ast$, is found
  for ${\mathcal{F}} = 1.0$.  The shaded are shows the region where
  $\Lambda^\ast$ falls by 1.0, or the $1 \sigma$ confidence 
interval for ${\mathcal{F}}$ [3].}
\label{scale_factor}
\end{center}
\end{figure}

\begin{figure}[t]
\begin{center}
\includegraphics[height=8.6cm,width=8.6cm]{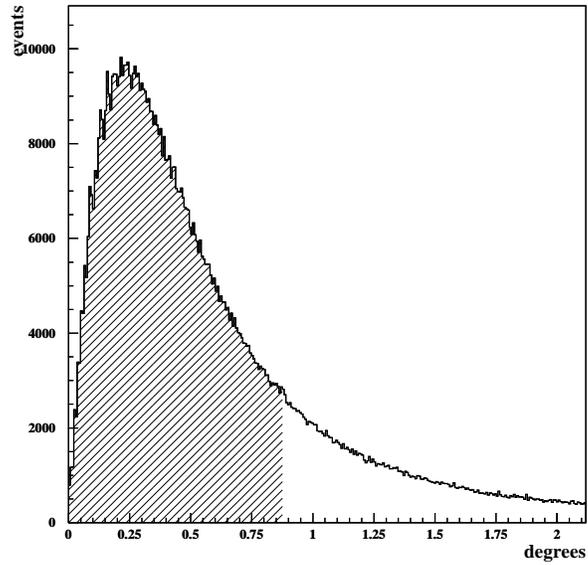}
\caption{The space angle distribution of double muons.  
The angular resolution of the MACRO apparatus, defined 
  as the cone angle containing 68\% of the events from a
  point source, is shown as a shaded region.  With this 
definition, MACRO's angular resolution is $ 0.90^\circ$.}
\label{ang_res}
\end{center}
\end{figure}

The distribution of $\Lambda$ 
as a function of $\mathcal{F}$ is shown
in Figure~\ref{scale_factor}.  The maximum $\Lambda$, $\Lambda^\ast$
is found for ${\mathcal{F}} = 1.0$ which implies
that the unscaled space angle distribution of double muons
should be used to determine the anglular resolution
of the MACRO apparatus.  
Many definitions have 
been used for the angular resolution (\cite{cygnus}, \cite{casa},
\cite{tibet}).  We choose the cone of 
angle $\theta_{68\%}$ that contains the 68$\%$ of the events from a
point-like source.  In Figure~\ref{ang_res} we show the 
space angle distribution of double muons.  
Using our definition of the angular resolution, Figure~\ref{ang_res} 
gives $\theta_{68\%} =
0.90^{\circ}$.  The $1 \sigma$ error limits on the angular
resolution can be estimated from the interval of $\mathcal{F}$ in which
$\Lambda^\ast$ falls by 1.0 \cite{PDG}.  The shaded region in
Figure~\ref{scale_factor} shows this interval to be ${\mathcal{F}} =
1.0 \pm 0.35$.  Using these values to scale the double muon distribution, 
we find that the angular resolution is 
$\theta_{68\%} = 0.9^\circ \pm 0.3^\circ$.

In Figure~\ref{moondens} we show as a solid line the expected event density 
in the moon window for an angular resolution of the MACRO apparatus 
of $0.90^\circ$.  The model used in this computation is given by  

\begin{equation}
  \frac{\Delta N }{\Delta \Omega} = 
     \Delta [ N^{bkd} - 
     S_M \cdot \tilde {\mathcal{P}} (\alpha; {\mathcal{F}} = 1.0) ] 
     / \Delta \Omega , 
\end{equation}

\medskip

\noindent where $\alpha = \sqrt{(x_s-x_i)^2 + (y_s-y_i)^2}$ 
is the angular distance from the moon center.  The
model fits the data well ($\chi^2 = 48.7/47$ d.o.f.).

\section{Conclusions}

The MACRO detector at a mean depth of 3700~m.w.e, operational since
February~1989, has collected a muon sample of about 39~million events.
Using this sample we have searched for the moon shadow cast in the cosmic
ray sky at primary energies $\sim 10-15$~TeV.  In the deficit  
analysis, 
we find an event deficit 
around the moon of significance 
$3.5 \sigma$.  With a maximum likelihood analysis, we confirm 
the detection of the moon's shadow 
with significance of $3.6 \sigma$.  
This is the first detection 
of the moon shadow underground.  Our estimate of the 
angular resolution is 
$\theta_{68\%} = 0.9^\circ \pm 0.3^\circ$.

These results demonstrate MACRO's capabilities as a muon telescope by
confirming its absolute pointing ability and quantifying its angular
resolution.  This investigation shows that the MACRO detector has the
capability of detecting signals from cosmic sources by observing secondary
cosmic muons underground.

\newpage

\newpage

\newpage

\newpage

\newpage

\newpage

\newpage

\newpage

\end{document}